\documentclass[a4paper, numberwithinsect, pdfa, UKenglish,cleveref, autoref, thm-restate]{socg-lipics-v2021}

\usepackage{algorithmicx}
\usepackage{algorithm}
\usepackage[noend]{algpseudocode}

\usepackage{tikz}
\usetikzlibrary{shapes.geometric}
\usetikzlibrary{arrows.meta}
\usepackage{indentfirst}

\usetikzlibrary{shadows}
\usetikzlibrary{decorations.pathreplacing,calligraphy,backgrounds}

\usetikzlibrary{3d, calc}
\usepackage{tikz-3dplot}

\pdfoutput=1 

\newcommand{\br}[1]{\mathopen{}\left( #1 \right)}
\newcommand{\brc}[1]{\mathopen{}\left\{ #1 \right\}}
\newcommand{\spr}[1]{\mathopen{}\left| #1 \right|}
\newcommand{\fl}[1]{\mathopen{}\left\lfloor #1 \right\rfloor}
\newcommand{\cl}[1]{\mathopen{}\left\lceil #1 \right\rceil}
\newcommand{\angl}[1]{\mathopen{}\langle #1 \rangle}

\newcommand{\diam}{\operatorname{diam}}

\newcommand{\OPT}{\texttt{OPT}}
\newcommand{\COST}{\texttt{COST}}
\newcommand{\argmin}{\operatorname*{argmin}}
\newcommand{\argmax}{\operatorname*{argmax}}

\newcommand{\vertex}[5]{%
  \begin{scope}[canvas is xy plane at z=#3]
    \node[circle, draw=#5, thick, inner sep=1.5pt, transform shape] (#1) at (#2) {#4};
  \end{scope}
}

\title{Searching in trees with $k$-up-modular cost functions} 

\titlerunning{Searching in trees with $k$-up-modular  functions} 

\author{Michał {Szyfelbein}}{Faculty of Electronics, Telecommunications and Informatics, Gdańsk University of Technology, Poland \and  }{s193307@student.pg.edu.pl}{https://orcid.org/0009-0009-9894-9671}{}

\authorrunning{M. Szyfelbein} 

\Copyright{Michał Szyfelbein} 



\category{} 

\relatedversion{} 

\acknowledgements{The author wants to thank Dariusz Dereniowski for the preliminary discussions and guidance}

\EventEditors{}
\EventLongTitle{}
\EventShortTitle{}
\EventAcronym{}
\EventYear{}
\EventDate{}
\EventLocation{}
\EventLogo{}
\SeriesVolume{}
\ArticleNo{}

\begin{document}

\maketitle

\begin{abstract}

Consider the following generalization of the classic binary search problem: a searcher is required to find a hidden vertex $x$ in a tree $T$. To do so, they iteratively perform queries to an oracle, each about a chosen vertex $v$. After each such call, the oracle responds whether the target was found and if not, the searcher receives as a reply the connected component of $T-v$ which contains $x$. Additionally, each vertex $v$ may have a different query cost $c(v)$. The goal is to find the optimal querying strategy which minimizes the worst case cost required to find $x$. The problem is known to be NP-hard even in restricted classes of trees such as bounded diameter spiders [Cicalese et al. 2016], and no constant factor approximation algorithm is known for general trees. Following the recent studies of [Dereniowski et al. 2022, Dereniowski et al. 2024], instead of restricted classes of trees, we explore restrictions on the cost function. We generalize the notion of up-monotonic functions and introduce the concept of \textit{$k$-up-modularity}. We show that an $O(\log\log n)$-approximate solution can be found within $k^{O(\log k)}\cdot\text{poly}(n)$ time.  
\end{abstract}
\newpage

\section{Introduction}
Searching plays a central role in computer science, forming the basis of numerous theoretical models and practical applications. Many fundamental computational problems, such as sorting, can be viewed as special cases of searching or rely on efficient search procedures as core subroutines. The design of search strategies is therefore of broad significance, with applications ranging from data management systems to artificial intelligence and operations research.  

A classical example is the \textit{binary search}, which is a strategy of searching, used to locate a target value within a linearly ordered set using a logarithmic amount of comparison operations. A natural application of binary search arises in experimental settings, where each such operation corresponds to performing a measurement that determines whether the target value is smaller or greater than a chosen threshold. This problem can be equivalently formulated as searching within a path: each comparison is modeled as a query to an oracle that, in constant time, verifies whether the queried vertex is the target and, if not, identifies the subpath which contains it.

\subsection{Problem description}
\begin{figure}[htbp]
    \begin{minipage}[t]{0.47\textwidth}
        \centering
        \begin{tikzpicture}[every node/.style={draw=none, very thick}, every path/.style={thick}]
        
        \node[label={[yshift=-2pt]above:{$1/5$}}] (1) at (-1,10) {$v_1$};
        \node[label={[yshift=-2pt]above:{$2/5$}}] (2) at (-3,8.85) {$v_2$};
        \node[label={[yshift=+2pt]below:{$1/5$}}] (3) at (-1.5,8.85) {$v_3$};
        \node[label={[yshift=-2pt]above:{$1/5$}}] (4) at (0,8.85) {$v_4$};
        \node[label={[yshift=+2pt]below:{$3/5$}}] (5) at (-4,7.7) {$v_5$};
        \node[label={[yshift=+2pt]below:{$4/5$}}] (6) at (-2,7.7) {$v_6$};
        \node[label={[yshift=+2pt]below:{$1$}}] (7) at (-0.5,7.7) {$v_7$};
        \node[label={[yshift=+2pt]below:{$2/5$}}] (8) at (1,7.7) {$v_8$};
        \node[label={[yshift=+2pt]below:{$3/5$}}] (9) at (0.5,6.55) {$v_9$};
        \node[label={[yshift=+2pt]below:{$1/5$}}] (10) at (-0.5,5.4) {$v_{10}$};
        \node[label={[yshift=+2pt]below:{$4/5$}}] (11) at (0.5,4.25) {$v_{11}$};
        
        \draw[] (1) to (2);
        \draw[] (1) to (3);
        \draw[] (1) to (4);
        \draw[] (2) to (5);
        \draw[] (2) to (6);
        \draw[] (4) to (7);
        \draw[] (4) to (8);
        \draw[] (7) to (9);
        \draw[] (9) to (10);
        \draw[] (10) to (11);
        
        \end{tikzpicture}
        \caption{Example input tree $T$ and cost function $c$.}\label{exampleInputTree}
    \end{minipage}%
    \hspace{0.06\textwidth}
    \begin{minipage}[t]{0.47\textwidth}
        \centering
        \begin{tikzpicture}[every node/.style={draw=none, very thick}, every path/.style={thick}]
        
        \node[label=above:{$1/5$}] (1) at (0,10) {$v_4$};
        \node[label=above:{$3/5$}] (2) at (-1,8.5) {$v_5$};
        \node[label=below:{$2/5$}] (3) at (0.5,8.5) {$v_8$};
        \node[label=above:{$3/5$}] (4) at (1.5,8.5) {$v_{9}$};
        \node[label=above:{$1/5$}] (5) at (-2,7) {$v_1$};
        \node[label=above:{$1$}] (6) at (1,7) {$v_7$};
        \node[label=above:{$1/5$}] (7) at (2.5,7) {$v_{11}$};
        \node[label=above:{$2/5$}] (8) at (-3,5.5) {$v_2$};
        \node[label=above:{$1/5$}] (9) at (-1,5.5) {$v_3$};
        \node[label=above:{$1/5$}] (10) at (1.5,5.5) {$v_{10}$};
        \node[label=above:{$4/5$}] (11) at (-2,4) {$v_6$};
        
        \draw[->] (1) to (2);
        \draw[->] (1) to (3);
        \draw[->] (1) to (4);
        \draw[->] (2) to (5);
        \draw[->] (4) to (6);
        \draw[->] (4) to (7);
        \draw[->] (5) to (8);
        \draw[->] (5) to (9);
        \draw[->] (7) to (10);
        \draw[->] (8) to (11);
        
        \end{tikzpicture}
        \caption{Example decision tree $D$ for $T$ of cost $11/5$.}\label{exampleDecisionTree}
    \end{minipage}
\end{figure}

In this work, we explore a generalization of the binary search problem by extending it to trees with non-uniform query costs. Specifically, we consider a tree $T$ and a cost function $c:V\to \mathbb{R}^+$, where $c\br{v}$ represents the cost of querying vertex $v$ (see Figure~\ref{exampleInputTree}). This cost can be interpreted as the amount of time units after which the response is returned. A search strategy is modeled as a decision tree $D$ (see Figure~\ref{exampleDecisionTree}), whose goal is to locate a hidden target vertex $x$ by issuing queries to vertices of $T$. A query to vertex $v$ incurs cost $c(v)$ and returns one of two possible responses: either confirms that $v$ is the target, thus, terminating the search, or identifies the connected component of $T-v$ which contains $x$. The quality of a decision tree is measured by its worst-case search time, i.e., the maximum total query cost required to identify the target. We wish to find a decision tree which minimizes the above cost.

The problem is known to be NP-hard for several restricted classes of trees \cite{DereniowskiVxRankOfChGsAndWTs, Dereniowski2009ERankOfWTs, Cicalese2012BinIdentPForWTs, Cicalese2016OnTSPwNonUniCost}. Moreover, no constant-factor approximation algorithm has been found, and the best currently known result achieves an $O\br{\sqrt{\log n}}$-approximation \cite{dereniowski2017ApproxSsForGeneralBSinWTs}. On the other hand, some progress has been made for cost functions with superimposed structure: constant-factor approximation algorithms are known for both down-monotonic and up-monotonic cost functions \cite{dereniowski2022CFApproxAlgForBSInTsWithMonoQTimes, dereniowski2024SInTsMonoQTs}. Motivated by this progress, we introduce a concept of $k$-up-modular cost functions, which generalizes the notion of up-monotonicity. For this setting, we develop an algorithm that achieves an $O(\log\log n)$-approximation in time $k^{O(\log k)}\cdot \text{poly}(n)$.

\subsection{Motivation and applications}

Searching in trees can be used to model various real-life applications. These include: scheduling of parallel database join operations \cite{DereniowskiEfPQProcByGRank,OnMinERSTs,MinERSTrofTGs}, automated bug detection in computer code \cite{OptimalSinT,dereniowski2022CFApproxAlgForBSInTsWithMonoQTimes,dereniowski2024SInTsMonoQTs}, parallel Cholesky factorization of matrices \cite{Dereniowski2003CholeskyFactofMx}, VLSI-layouts \cite{OnAGPartWAppVLSI}, hierarchical clustering of data \cite{Acostfunctionforsimilaritybasedhierarchicalclustering,HCObjFsandAlgs,Approximatehierarchicalclusteringviasparsestcutandspreadingmetrics} and parallel assembly of multi-part products from their components \cite{ParAofModPs,Dereniowski2009ERankOfWTs}.

The study of $k$-up-modular functions is primarily motivated by their potential applications in the automatic testing of large, distributed systems. In this setting, each vertex corresponds to a given block of code, while each child of a vertex represents a smaller subblock of the same code. The code is assumed to contain a defect, which is required to be found. Each query represents an automated test that verifies whether a given block contains the bug. A natural assumption is that the cost of a query is highest for a vertex representing the entire system and gradually decreases, as we move towards its smaller subcomponents. Such a cost function is called up-monotonic and was analyzed in \cite{dereniowski2022CFApproxAlgForBSInTsWithMonoQTimes, dereniowski2024SInTsMonoQTs}.

However, in practice, testing a large distributed system often involves replacing certain components with mock objects, i.e., artificially constructed substitutes. From the perspective of the test, a mock behaves like the original component in that it produces identical outputs. In reality, however, it merely returns precomputed values required for the test to proceed. This practice significantly reduces the runtime of tests and ensures the reproducibility of their outcomes. For instance, a mocked component may represent a large database or an external service running on a remote, potentially slower machine. In such cases, testing a component in isolation may be more costly than testing a larger system module that uses its mock. As a result, the query cost function may no longer be strictly up-monotonic. To capture such irregularities, we introduce the concept of $k$-up-modularity and show how to obtain an efficient search strategy in FPT time with respect to $k$.

From the theoretical point of view, the motivation of such notion is to find a middle ground between: the constant-factor approximation known for a very narrow class of monotonic functions \cite{dereniowski2022CFApproxAlgForBSInTsWithMonoQTimes,dereniowski2024SInTsMonoQTs}, and $O\br{\sqrt{\log n}}$-approximation known for the general case \cite{dereniowski2017ApproxSsForGeneralBSinWTs}. This is also visible in the structure of our solution. To construct the strategy, we process consecutive intervals of costs recursively, following an approach similar to the one used for monotonic cost functions  \cite{dereniowski2022CFApproxAlgForBSInTsWithMonoQTimes, dereniowski2024SInTsMonoQTs}. However, more general cost function require a novel way of extending the solution to subsequent intervals. Our algorithm builds the strategy in a top-down manner, employing the QPTAS designed in \cite{dereniowski2017ApproxSsForGeneralBSinWTs} and exploiting the connection of searching to the vertex ranking labeling \cite{OnakParys2006GenOfBSSInTsAndFLikePosets, Mozes_Onak2008FindOptTSStartInLinTime, Schaffer1989OptNodeRankOfTsInLinTime}. By combining the two, we are able to schedule all queries with costs within the currently processed interval. The procedure is then applied recursively until a full decision tree is obtained. Crucially, we show that the depth of this recursion is of order $O\br{\log \log n}$, which yields the desired guarantee on the quality of the solution.

\subsection{State-of-the-art and related results}
\paragraph*{Vertex Query Model}
For trees with uniform query costs, the optimal search strategy can be computed in linear time, as shown independently in \cite{OnakParys2006GenOfBSSInTsAndFLikePosets} and \cite{Schaffer1989OptNodeRankOfTsInLinTime}. The variant with non-uniform query costs becomes strongly NP-hard, but it is FPT (fixed-parameter tractable) with respect to the maximum cost \cite{DereniowskiVxRankOfChGsAndWTs}. For paths with non-uniform costs, a dynamic programming algorithm with running time $O(n^2)$ exists \cite{Cicalese2012BinIdentPForWTs, LaberOnBSWithNonUniCosts}. For general trees, a polynomial-time $O(\sqrt{\log n})$-approximation is known \cite{dereniowski2017ApproxSsForGeneralBSinWTs}, derived from a QPTAS that achieves a $(1+\varepsilon)$-approximation in $n^{O(\log n / \varepsilon^2)}$ time. For monotonic cost functions, constant-factor approximation algorithms have been developed \cite{dereniowski2022CFApproxAlgForBSInTsWithMonoQTimes, dereniowski2024SInTsMonoQTs}. Related versions of the problem include the average-case criterion, both for unitary costs \cite{Angelidakis2018ShortestPQ, Fast_app_centroid_trees, SplayTonT,Berendsohn2024, OnDasHC, ATightAnalysisofGreedy} and for non-uniform costs setup \cite{Cicalese2016DecTreesSimEval, Acostfunctionforsimilaritybasedhierarchicalclustering}.

\paragraph*{Edge Query Model}
The uniform-cost, search problem in trees with edge queries requires intricate algorithms, despite admitting linear-time solutions \cite{Lam1998ERankOfGsIsH, Mozes_Onak2008FindOptTSStartInLinTime}. Upper bounds on the depth of optimal decision trees were studied in \cite{LaberFastSInTs, MAKINOOnMinERankSTs, DereniowskiEfPQProcByGRank}, with the best currently known bound stating that at most $\tfrac{\Delta\br{T}-1}{\log\br{\Delta\br{T}+1}-1}\cdot \log n$ queries always suffice \cite{Emamjomeh2016DetAndProbBSinGs}.
 For non-uniform costs, the problem becomes NP-hard, even for restricted classes of trees: trees of degree $\Delta\br{T}\geq 3$ \cite{Cicalese2012BinIdentPForWTs} and spiders with diameter $\diam\br{T}\geq 6$ \cite{Cicalese2016OnTSPwNonUniCost}. For paths and trees with diameter $\diam\br{T}\leq 5$, polynomial-time solutions exist \cite{Cicalese2012BinIdentPForWTs}. For general trees, known approximation algorithms range from $O(\log n)$ approximation in  \cite{Dereniowski2009ERankOfWTs}, to $O(\log n / \log \log \log n)$ \cite{Cicalese2012BinIdentPForWTs}, further improved to $O(\log n / \log \log n)$ \cite{Cicalese2016OnTSPwNonUniCost} and to $O\br{\sqrt{\log n}}$, by a reduction from the vertex query version of the problem \cite{dereniowski2017ApproxSsForGeneralBSinWTs}. Average-case formulations have also been studied for both uniform \cite{Acostfunctionforsimilaritybasedhierarchicalclustering, OnDasHC, ATightAnalysisofGreedy, Approximatehierarchicalclusteringviasparsestcutandspreadingmetrics, HCObjFsandAlgs}, and non-uniform query costs \cite{Jacobs2010OnTheComplexSearchInTsAvg, Cicalese2014ImprovedApproxAvgTs, Hgemo2024TightAB, Approximatehierarchicalclusteringviasparsestcutandspreadingmetrics, HCObjFsandAlgs}.


\subsection{Organization of the paper}
In Section \ref{preliminaries}, we introduce the necessary preliminaries (Section \ref{notationAndQueryModel}) and formal definitions used throughout the paper, including decision trees and their cost (Section \ref{definitionOfDecisionTree}), as well as heavy modules and $k$-up modular functions (Section \ref{kUpModularity}).

Section \ref{parametrizedSolution} contains the main result: an $O\br{\log\log n}$-approximation algorithm running in $k^{O\br{\log k}}\cdot\text{poly}\br{n}$ time. To obtain this result, we first introduce the notion of \emph{cost levels} (Section \ref{costLevels}). Next, in Section \ref{vertexRanking}, we highlight the correspondence between \emph{vertex ranking labeling} and searching in trees, which enables efficient processing of (almost) uniform-cost instances.

The core of our approach is presented in Section \ref{mainRecursiveProcedure}, where we describe a recursive procedure that, given a tree $\mathcal{T}$, a cost function $c$ and an interval $\left(a,b\right]$, schedules all queries to heavy queries, i.e. vertices $v\in V\br{\mathcal{T}}$, with $c\br{v}\in \left(a,b\right]$. This is achieved by identifying a set $\mathcal{Z}\subseteq V\br{\mathcal{T}}$, such that each $\mathcal{T}'\in \mathcal{T}-\mathcal{Z}$ contains at most one connected component with heavy vertices. We then construct an \emph{auxiliary tree} $\mathcal{T}_{\mathcal{Z}}$ and use the QPTAS from \cite{dereniowski2017ApproxSsForGeneralBSinWTs} to find a decision tree $D_{\mathcal{Z}}$ for $\mathcal{T}_{\mathcal{Z}}$. The details of this construction are provided in Section \ref{auxTreeConstruction}. Then,  for each $\mathcal{T}'\in \mathcal{T}-\mathcal{Z}$, we build a decision tree $D_H$ containing queries to all of its heavy vertices using the connection of searching with the vertex ranking labeling. After that, we attach $D_H$ below an appropriate query in $D_{\mathcal{Z}}$. Once all heavy queries are scheduled, the rest of the decision tree is build by applying the procedure recursively.

In Section \ref{conslusionsAndFutureWork} we summarize the results, provide the conclusions and outline possible directions for future research.

\section{Preliminaries}\label{preliminaries}
\subsection{Notation and query model}\label{notationAndQueryModel}

For any tree $T$, by $uv\in E\br{T}$ we denote an edge connecting vertices $u$ and $v$ in $T$. If a given tree $T$ is rooted, we denote its root as $r\br{T}$. For any $v\in V\br{T}$, by $T-v$ we denote the set of all connected components occurring after deleting $v$ from $T$. For any $S\subseteq V\br{T}$, by $T[S]$ we denote the forest induced by $S$, by $T\angl{S}$ the minimal connected subtree of $T$ containing all vertices from $S$, and by $T-S$ the set of all connected components occurring after deleting vertices in $S$ from $T$.

The set of neighbors of $v\in V\br{T}$ will be denoted as $N_T\br{v} = \brc{u\in V\br{T}|uv\in E\br{T}}$, and the set of neighbors of a subtree $T'\subseteq T$ as $N_T\br{T'} = \bigcup_{v\in V\br{T'}}N_T\br{v}-V\br{T'}$. Let $\deg_{T}\br{v}=\spr{N_T\br{v}}$ denote the degree of $v$ in $T$. By $\mathcal{P}_{T}\br{u, v}=T\angl{\brc{u,v}}-\brc{u,v}$ we denote the path of vertices between $u$ and $v$ in $T$ (excluding $u$ and $v$). Analogously, for $V_1,V_2\subseteq V\br{T}$ we define $\mathcal{P}_{T}\br{V_1, V_2}=T\angl{V_1\cup V_2}-\br{V_1\cup V_2}$.

The \textit{Tree Search Instance} consists of a pair $\br{T,c}$, where $c:V\br{T}\to \mathbb{R}_{+}$ is the cost of querying each vertex $v$ (often interpreted as the time required to query $v$). During the \textit{Search Process}, the searcher is allowed to iteratively perform \textit{queries}, each of which asks about a chosen vertex $v$ and returns the answer after time $c\br{v}$. If the answer is affirmative, then $v$ is the target, otherwise, a connected component $T'\in T-x$ is returned, such that $x\in V\br{T'}$. Based on this information, the searcher narrows the subtree of $T$ which contains $x$. We call this subtree the \textit{candidate} subtree and its vertices the \textit{candidate} set. This process continues until the last vertex is queried and the position of the target is revealed\footnote{One may distinguish two distinct search models here. In the first, the process continues until the candidate set becomes empty, i.e., the target itself must eventually be queried. In the second, if all neighbors of $x$ have already been queried, the query to $x$ can be omitted. Although the models are not identical, their optima differ by at most a factor of 2, which does not affect the asymptotic quality of our solution.}

\subsection{Definition of a decision tree}\label{definitionOfDecisionTree}
A decision tree for $T$ is a rooted tree $D=(V\br{D}, E\br{D})$, where $V\br{D}=V\br{T}$ are the vertices of $D$ and $E\br{D}$ are the edges of $D$. It is required that each child of $q\in V\br{D}$ corresponds to a distinct response to the query at $q$, with respect to the subtree of candidate solutions that remain after performing all previous queries. 

Let $Q_D\br{T,x}$ denote the sequence of queries made to locate a target $x \in V\br{T}$ using $D$, i. e., the sequence of vertices belonging to the unique path in $D$ starting at $r\br{D}$ and ending at $x$. We define the worst-case cost of a decision tree $D$ in $\br{T,c}$ as:
$$
\COST_D\br{T, c} = \max_{x\in V\br{T}} \brc{\sum_{q\in Q_{D}\br{T,x}}c\br{q }}.
$$

We also define the (optimal) cost of $\br{T,c}$ as:
$$
\OPT\br{T, c} = \min_{\text{decision tree }D} \brc{\COST_D\br{T, c}}.
$$

A given decision tree is \textit{optimal} if its cost is equal to $\OPT\br{T, c}$. In the \textit{Tree Search Problem}, one is required to find the optimal decision tree $D$ for the instance $\br{T,c}$. Note that, whenever the context is clear, we will omit the weight function and write simply $\OPT(T)$ and $\COST_D(T)$ for brevity.

Let $D'$ be a subtree of a valid decision tree $D$ for $T$ containing $r\br{D}$. We will say that $D'$ is a \textit{partial} decision tree for $D$, and we define its cost analogously to that of $D$ (although we only count queries belonging to $D'$). 

From now on, we will assume that all costs are normalized, so that $\max_{v\in V\br{T}}\brc{c\br{v}}=1$. If not, the costs are scaled by dividing them by $\max_{v\in V\br{T}}\brc{c(v)}$. Note that this operation does not affect the optimality of a strategy or the quality of an approximation. Following this, an observation stated in \cite{dereniowski2017ApproxSsForGeneralBSinWTs} is easily obtained:
\begin{observation}\label{basicBoundsOnCost}
    Let $T$ be a tree and $c:V\to \mathbb{R}^+$, such that $\max_{v\in V\br{T}}\brc{c\br{v}}=1$ be a normalized cost function. Then, $1\leq \OPT\br{T, c} \leq \fl{\log n}+1$.
\end{observation}

\subsection{$k$-up-modularity}\label{kUpModularity}
The main algorithmic difficulty in dealing with the problem arises when the values of the cost function vary drastically. We would like to measure this "irregularity" in a quantifiable way. To do so, we introduce the notion of $k$-up-modularity.

\begin{figure}[htbp]
\centering
\usetikzlibrary{3d}
\usetikzlibrary{calc}

\def\angi{27}
\def\angii{62}
\pgfmathsetmacro\xx{sin(\angii)}
\pgfmathsetmacro\xy{-cos(\angii)*sin(\angi)}
\pgfmathsetmacro\yx{sin(\angii-90)}
\pgfmathsetmacro\yy{-cos(\angii-90)*sin(\angi)}
\pgfmathsetmacro\zx{0}
\pgfmathsetmacro\zy{cos(\angi)}

\begin{tikzpicture}[ scale=2,
  x={({\xx cm,\xy cm})},
  y={({\yx cm,\yy cm})},
  z={({\zx cm,\zy cm})},
  line cap=round, line join=round]

\begin{scope}[canvas is xy plane at z=1.05]
    \fill[yellow!20!orange!20, opacity=0.5] (-0.1,-0.1) rectangle (5.5,3.8);
    \draw[yellow!50!orange, very thick] (-0.1,-0.1) rectangle (5.5,3.8);
    \node[anchor=north west, font=\small\bfseries, yellow!40!orange] at (0.1,0.1) {$s$};
  \end{scope}
  \draw[->, very thick] (0,0,0) -- (5.5,0,0) node[anchor=north east] {};
  \draw[->, very thick] (0,0,0) -- (0,4,0) node[anchor=north west] {};
  \draw[->, very thick] (0,0,0) -- (0,0,1.75) node[anchor=south] {$c\br{v}$};

    \vertex{1}{0.25,0.25}{0.0}{}{};
    \vertex{2}{0.7,0.45}{0.0}{}{};
    \vertex{3}{1.25,0.75}{0.0}{}{};
    \vertex{4}{1.7,1.15}{0.0}{}{};
    
    \vertex{5}{1.55,1.7}{0.0}{}{};
    \vertex{6}{1.5,2.3}{0.0}{}{};
    \vertex{7}{1.4,3}{0.0}{}{};
    \vertex{8}{0.9,3.3}{0.0}{}{};
    \vertex{9}{0.3,3.5}{0.0}{}{};
    
    \vertex{10}{2.75,0.75}{0.0}{}{};
    \vertex{11}{2.75,1.4}{0.0}{}{};
    \vertex{12}{2.5, 1.75}{0.0}{}{};
    
    \vertex{13}{3.2,0.6}{0.0}{}{};
    \vertex{14}{3.75,0.6}{0.0}{}{};
    \vertex{15}{4.2,0.5}{0.0}{}{};
    \vertex{16}{5,0.4}{0.0}{}{};

  \draw[very thick] (1) to (2);
  \draw[very thick] (2) to (3);
  \draw[very thick] (3) to (4);
  \draw[very thick] (4) to (5);
  \draw[very thick] (5) to (6);
  \draw[very thick] (6) to (7);
  \draw[very thick] (7) to (8);
  \draw[very thick] (8) to (9);
  
  \draw[very thick] (4) to (10);
  \draw[very thick] (10) to (11);
  \draw[very thick] (11) to (12);
  
  \draw[very thick] (10) to (13);
  \draw[very thick] (13) to (14);
  \draw[very thick] (14) to (15);
  \draw[very thick] (15) to (16);

\vertex{91}{0.25,0.25}{0.35}{}{green!50!black, dotted};
\vertex{92}{0.7,0.45}{0.9}{}{green!50!black, dotted};
\vertex{93}{1.25,0.75}{1.75}{}{blue};
\vertex{94}{1.7,1.15}{1.5}{}{blue};

\vertex{95}{1.55,1.7}{1.25}{}{blue};
\vertex{96}{1.5,2.3}{0.95}{}{green!50!black, dotted};
\vertex{97}{1.4,3}{0.8}{}{green!50!black, dotted};
\vertex{98}{0.9,3.3}{1.65}{}{blue};
\vertex{99}{0.3,3.5}{1.25}{}{blue};

\vertex{910}{2.75,0.75}{1.75}{}{blue};
\vertex{911}{2.75,1.4}{1.65}{}{blue};
\vertex{912}{2.5, 1.75}{0.8}{}{green!50!black, dotted};

\vertex{913}{3.2,0.6}{1.65}{}{blue};
\vertex{914}{3.75,0.6}{1.35}{}{blue};
\vertex{915}{4.2,0.5}{0.95}{}{green!50!black, dotted};
\vertex{916}{5,0.4}{0.7}{}{green!50!black, dotted};

  \draw[->, green!50!black, dotted, dashed] (1) to (91);
  \draw[->, green!50!black, dotted, dashed] (2) to (92);
  \draw[green!50!black, dotted, dashed] (3) to (1.25,0.75, 1.05);
  \draw[->, blue, dashed] (1.25,0.75, 1.05) to (93);
  
  \draw[green!50!black, dotted, dashed] (4) to (1.7,1.15, 1.05);
  \draw[->, blue, dashed] (1.7,1.15, 1.05) to (94);

  \draw[green!50!black, dotted, dashed] (5) to (1.55,1.7, 1.05);
  \draw[->, blue, dashed] (1.55,1.7, 1.05) to (95);
  
  \draw[blue, dashed] (6) to (1.5,2.3, 0.19);
  \draw[->, green!50!black, dotted, dashed] (1.5,2.3, 0.19) to (96);
  
  \draw[blue, dashed] (7) to (1.4,3, 0.6);
  \draw[->, green!50!black, dotted, dashed] (1.4,3, 0.6) to (97);
  
  \draw[blue, dashed] (8) to (0.9,3.3, 0.77);
  \draw[green!50!black, dotted, dashed] (0.9,3.3, 0.77) to (0.9,3.3, 1.05);
  \draw[->, blue, dashed] (0.9,3.3, 1.05) to (98);

  \draw[blue, dashed] (9) to (0.3,3.5, 0.88);
  \draw[green!50!black, dotted, dashed] (0.3,3.5, 0.88) to (0.3,3.5, 1.05);
  \draw[->, blue, dashed] (0.3,3.5, 1.05) to (99);

  \draw[green!50!black, dotted, dashed] (10) to (2.75,0.75, 1.05);
  \draw[->, blue, dashed] (2.75,0.75, 1.05) to (910);
  ;
  
  \draw[green!50!black, dotted, dashed] (11) to (2.75,1.4, 1.05);
  \draw[->, blue, dashed] (2.75,1.4, 1.05) to (911);
  
  \draw[->, green!50!black, dotted, dashed] (12) to (912);
  
  \draw[green!50!black, dotted, dashed] (13) to (3.2,0.6, 1.05);
  \draw[->, blue, dashed] (3.2,0.6, 1.05) to (913);

  \draw[green!50!black, dotted, dashed] (14) to (3.75,0.6, 1.05);
  \draw[->, blue, dashed] (3.75,0.6, 1.05) to (914);
  
  \draw[->, green!50!black, dotted, dashed] (15) to (915);

  \draw[blue, dashed] (16) to (5,0.4, 0.51);
  \draw[->, green!50!black, dotted , dashed] (5,0.4, 0.51) to (916);

    \draw[very thick, green!50!black, dotted] (91) to (92);
  
  \draw[very thick, green!50!black, dotted] (92) to (0.8,0.5, 1.05);
  \draw[very thick, blue] (0.8,0.5, 1.05) to (93);
  
  \draw[very thick, blue] (93) to (94);
  \draw[very thick, blue] (94) to (95);
  
  \draw[very thick, blue] (95) to (1.55,2.25, 1.05);
  \draw[very thick, green!50!black, dotted] (1.55,2.25, 1.05) to (96);
  
  \draw[very thick, green!50!black, dotted] (96) to (97);

  \draw[very thick, green!50!black, dotted] (97) to (1.23,3.07, 1.05);
  \draw[very thick, blue] (1.23,3.07, 1.05) to (98);
  
  \draw[very thick, blue] (98) to (99);
  
  \draw[very thick, blue] (94) to (910);
  \draw[very thick, blue] (910) to (911);
  
  \draw[very thick, blue] (911) to (2.59, 1.7, 1.05);
  \draw[very thick, green!50!black, dotted] (2.59, 1.7, 1.05) to (912);
  
  \draw[very thick, blue] (910) to (913);
  \draw[very thick, blue] (913) to (914);
  
  \draw[very thick, blue] (914) to (4.08,0.52, 1.05);
  \draw[very thick, green!50!black, dotted] (4.08,0.52, 1.05) to (915);
  
  \draw[very thick, green!50!black, dotted] (915) to (916);

\end{tikzpicture}
\caption{A visual depiction of a tree $T$ with a $3$-up-modular cost function $c$. Each vertex of a tree is mapped onto some value of $c$. The yellow plane represents some threshold value $t\in \mathbb{R}_{\geq 0}$ (in this particular example $k\br{T,c,t}=2$). The (two) blue subtrees represent members of $\mathcal{H}_{T,c}\br{t}$.}\label{kUpModularityExample}

\end{figure}

Let $t\in\mathbb{R}_{\geq0}$. We define a \textit{heavy module} with respect to $t$ as $H\subseteq V\br{T}$ such that, $T[H]$ is connected, for every $v \in H$, $c\br{v} > t$, and $H$ is maximal - no vertex can be added to it without violating one of its properties. We then define the \textit{heavy module set} with respect to $t$ in $\br{T,c}$ as:
$$
\mathcal{H}_{T,c}\br{t}=\brc{H\subseteq V\br{T}\mid H \text{ is a heavy module w.r.t. } t},
$$

Let $k\br{T,c, t} = \spr{\mathcal{H}_{T,c}\br{t}}$ be the size of the heavy module set, and finally let $k\br{T,c}= \max_{s\in\mathbb{R}_{\geq 0}}\brc{k\br{T,c, t}}$. We say that a function $c$ is $k$\textit{-up-modular} in $T$ when $k\geq k\br{T,c}$. Whenever clear from the context, we will use $k\br{T,c}$, $k\br{T}$, or $k$ to denote the lowest value such that $c$ is $k$-up-modular in $T$. To illustrate the notion of $k$-up-modularity, see Figure \ref{kUpModularityExample}.

The concept of $k$-up-modularity is a direct generalization of the notion of up-monotonicity of the cost function introduced in \cite{dereniowski2022CFApproxAlgForBSInTsWithMonoQTimes} (as monotonicity) and in \cite{dereniowski2024SInTsMonoQTs} (as up-monotonicity). Let $z=\argmax_{v\in V\br{T}}\brc{c\br{v}}$. A function $c$ is \textit{up-monotonic} in $T$ if for every $v,u\in V\br{T}$, whenever $v$ lies on the path between $z$ and $u$, we have $c\br{v}\geq c\br{u}$. 

It is easy to see that 1-up-modularity is equivalent to up-monotonicity. Observe that if $c$ is up-monotonic in $T$, then for every $t\in\mathbb{R}_{\geq 0}$, $T[V\br{T}-\brc{v\in V\br{T}\mid c\br{v}\leq t}]$ is connected and forms a single heavy module. Conversely, let $r=\argmax_{v\in V\br{T}}\brc{c\br{v}}$ and $u$ be any other vertex. If $c$ is 1-up-modular in $T$, then there is no vertex $v$ on the path between $r$ and $u$ such that $c\br{v}<c\br{u}$. Otherwise, for any $t\in \br{c\br{v},c\br{u}}$, $v$ does not belong to any heavy module, but $u$ and $r$ do. Since $v$ lies between them, $\spr{\mathcal{H}_{T,c}\br{t}}>1$, a contradiction.

\section{The parametrized $O\br{\log\log n}$-approx. solution}\label{parametrizedSolution}
\subsection{Cost levels}\label{costLevels}
The main idea of the algorithm is to partition vertices into intervals called \textit{cost levels} and process them in a top-down manner. At each level of the recursion, the algorithm schedules all necessary queries to vertices belonging to the given cost level. The rest of the decision tree is then built recursively. We consider the following intervals\footnote{We present the intervals in the ascending order in which a complete solution for each of them is obtained. However, since the procedure is recursive, the order in which the recursive calls are made is reverse.}:

\begin{enumerate}
    \item Firstly, an interval $\left( 0,{1}/{\log n}\right]$.
    \item Then, each subsequent interval $\mathcal{I}'=\left(a',b'\right]$ starts at the left endpoint of the previous interval $\mathcal{I}=\left(a,b\right]$, that is, $a'=b$, and ends with $b'=\min\brc{2b, 1}$. 
    
    This results in the following sequence of intervals, which partitions the interval $\left(0,1\right]$:
    
    $$\left( 0,{1}/{\log n}\right],\left({1}/{\log n},{2}/{\log n}\right], \left({2}/{\log n},{4}/{\log n}\right],\dots, \left({2^{\cl{\log\log n}-1}}/{\log n},1\right].$$
\end{enumerate}

We will ensure that when we call our procedure with parameters $\br{T, c, \left({2^{\cl{\log\log n}-1}}/{\log n},1\right]}$, the returned decision tree will be a valid decision tree for $T$.

We are now ready to introduce the notions of heavy and light vertices (and queries to them). We say that a vertex $v$ (or a query to it) is \textit{heavy} with respect to the interval $\mathcal{I}=\left(a,b\right]$ when $c\br{v}>a$. Otherwise, i.e., if $c\br{v}\leq a$, the vertex (and the query to it) is \textit{light} with respect to $\mathcal{I}$. Note that each heavy vertex belongs to some heavy module. Whenever clear from the context, we will omit the phrase "with respect to" and simply call the vertices and queries heavy and light.

\subsection{Vertex ranking}\label{vertexRanking}
The \textit{vertex ranking} of $T$ is a labeling of vertices $l:V\to \brc{1,2,\dots,\cl{\log n}}$, which satisfies the following condition: for each pair of vertices $u,v\in V\br{T}$, whenever $l\br{u}=l\br{v}$, there exists $z\in \mathcal{P}_T\br{u,v}$ for which $l\br{z}>l\br{v}$. Such a labeling always exists and can be computed in linear time by means of dynamic programming \cite{Schaffer1989OptNodeRankOfTsInLinTime, OnakParys2006GenOfBSSInTsAndFLikePosets, Mozes_Onak2008FindOptTSStartInLinTime}. 

Having a vertex ranking of $T$, one can easily obtain a decision tree for $T$ using the following procedure:
\begin{enumerate}
    \item Let $z\in V\br{T}$ be the unique vertex, such that for every $v\in V\br{T}$, $l\br{z}\geq l\br{v}$.
    \item Schedule a query to $z$ as the root of the decision tree $D$ for $T$.
    \item For each $T'\in T-z$, build a decision tree $D_{T'}$ recursively and hang it below the query to $z$ in $D$.
\end{enumerate} 

When the input tree has uniform costs and the ranking uses the minimal number of labels, the decision tree built in this way is optimal and never uses more than $\fl{\log n} + 1$ queries \cite{OnakParys2006GenOfBSSInTsAndFLikePosets}. Let \textsc{RankingBasedDT} be the name of the latter procedure. We have the following corollary:
  
\begin{corollary}\label{vertexRankingCorollary}
    There exists an $O\br{n}$ time procedure \textsc{RankingBasedDT} that finds the optimal decision tree for the Tree Search Problem when all costs are uniform. Moreover, the depth of such a decision tree, i.e., the worst-case number of queries, is at most $\fl{\log n} + 1$.
\end{corollary}

\subsection{The main recursive procedure}\label{mainRecursiveProcedure}

We are ready to present the main recursive procedure. To avoid ambiguity, let $\mathcal{T}$ be the subtree of $T$ processed at some level of the recursion. Alongside $\mathcal{T}$ and a cost function $c$, the algorithm takes as input an interval $\left(a,b\right]$, such that for every $v\in V\br{\mathcal{T}}$, $c\br{v}\leq b$ and $2a\geq b$. 
The basic steps of the Algorithm \ref{createDecisionTree} are as follows: 
\begin{enumerate}
    \item If every vertex is heavy, return a decision tree built by calling the \textsc{RankingBasedDT} procedure for $\mathcal{T}$.
    \item Otherwise, find a set $\mathcal{Z}$, such that each connected component of $\mathcal{T}'\in \mathcal{T}-\mathcal{Z}$ contains at most one heavy module.
    \item Create an auxiliary tree $T_{\mathcal{Z}}$ using the vertices of $\mathcal{Z}$ and create a new decision tree $D_{\mathcal{Z}}$ for $T_{\mathcal{Z}}$, using the QPTAS from \cite{dereniowski2017ApproxSsForGeneralBSinWTs}.
    \item For each $\mathcal{T}'\in \mathcal{T}-\mathcal{Z}$, build a decision tree $D_H$, by calling the \textsc{RankingBasedDT} procedure for $\mathcal{T}'\angl{H}$. Then, hang $D_H$ below the last query to $v\in N_{\mathcal{T}}\br{\mathcal{T}'}$ in $D_{\mathcal{Z}}$.
    \item For each $L\in\mathcal{T}'-H$, build a decision tree recursively. Then, hang $D_L$ below the last query to a vertex $v \in N_{\mathcal{T'}}\br{L}$ in $D_{\mathcal{Z}}$.
    \item Return the resulting decision tree $D$.
\end{enumerate}

Before providing a detailed description and analysis of the above procedure, we first present some basic properties necessary for the subsequent considerations. In particular, we will make use of the following well-known lemma \cite{Cicalese2016DecTreesSimEval}:
\begin{lemma}\label{subtreeOptLemma}
    Let $T'$ be a subtree of $T$. Then, $\OPT\br{T'}\leq\OPT\br{T}$.
\end{lemma}

\begin{algorithm}
\caption{The main recursive procedure}
\label{createDecisionTree}
\begin{algorithmic}[1]
    \Procedure{CreateDecisionTree}{$\mathcal{T},c,\left(a,b\right]$}
    \If{$b\leq 1/\log n$ or for every $v\in\mathcal{T}$, $c\br{v}>a$}\Comment{\texttt{Every $v\in\br{\mathcal{T}}$ is heavy}}
    \State \Return\label{basecaseDT} \Call{RankingBasedDT}{$\mathcal{T}$}. \Comment{\texttt{Apply Corollary \ref{vertexRankingCorollary}}.}
    \Else\Comment{\texttt{There are light vertices in $\mathcal{T}$.}}
        \State $\mathcal{X}\gets\emptyset$.
        \For{$H\in\mathcal{H}_{\mathcal{T}, c}\br{a}$} 
        \State Pick arbitrary $v\in H$ and add $v$ to $\mathcal{X}$.
        \EndFor
        \State $\mathcal{Z}\gets\mathcal{Y}\gets\mathcal{X}\cup \brc{v\in V\br{\mathcal{T}\angl{\mathcal{X}}}|\deg_{\mathcal{T}\angl{\mathcal{X}}}\br{v}\geq3}$.
        \For{$u,v\in \mathcal{Y},\mathcal{P}_{\mathcal{T}}\br{u, v}\neq\emptyset,\mathcal{P}_{\mathcal{T}}\br{u, v}\cap \mathcal{Y}=\emptyset$}
        \State Add $\argmin_{z\in \mathcal{P}_{\mathcal{T}}\br{u, v}}\brc{c\br{z}}$ to $\mathcal{Z}$. \Comment{\texttt{Lightest $v\in P_\mathcal{T}\br{u,v}$.}}
        \EndFor
        \State $\mathcal{T}_{\mathcal{Z}}=\br{\mathcal{Z}, \brc{uv|\mathcal{P}_{\mathcal{T}}\br{u, v}\cap \mathcal{Z}=\emptyset}}$.\Comment{\texttt{Build the auxiliary tree.}}
        \State $D\gets D_{\mathcal{Z}}\gets \Call{QPTAS}{\mathcal{T}_{\mathcal{Z}}, c, \epsilon =1}$.\label{QPTAScall}\Comment{\texttt{Apply Theorem \ref{QPTAS}.}}
        \For{$\mathcal{T}'\in \mathcal{T}-\mathcal{Z}$}
            \State $H\gets$ the unique heavy module in $\mathcal{T}'$.
            \State $D_{H}\gets$\Call{RankingBasedDT}{$\mathcal{T}'\angl{H}$}.\label{rankingbaseddtH}\Comment{\texttt{Apply Corollary \ref{vertexRankingCorollary}}.}
            \State Hang $D_H$ in $D$ below the last query to $v\in N_{\mathcal{T}}\br{\mathcal{T}'}$.\Comment{\texttt{By Obs. \ref{neighborsPathObservation}.}}
            \For{$L\in \mathcal{T}'-H$}
                \State $D_L\gets \Call{CreateDecisionTree}{L, c,\left(a/2,a\right]}$.\label{recursion}
                \State Hang $D_L$ in $D$ below the last query to $v\in N_{\mathcal{T}'}\br{L}$. \Comment{\texttt{By Obs. \ref{neighborsPathObservation}.}}
            \EndFor
        \EndFor
    \State \Return $D$
    \EndIf
    \EndProcedure
\end{algorithmic}
\end{algorithm}

We will also use the result of \cite{dereniowski2017ApproxSsForGeneralBSinWTs} regarding the existence of a QPTAS for the Tree Search Problem:

\begin{theorem}\label{QPTAS}
     For any fixed $0 < \epsilon \leq 1$, there exists a $(1+\epsilon)$-approximation algorithm for the Tree Search Problem running in $n^{O\br{{\log n}/{\epsilon^2}}}$ time.
\end{theorem}

For the rest of the analysis, fix $\mathcal{H}=\mathcal{H}_{\mathcal{T},c}\br{a}$ to be the set of heavy modules in $\mathcal{T}$. We have the following observations, which will be useful in the description and analysis of the algorithm:

\begin{observation}\label{heavymodulesetsize}
Let $\mathcal{H}$ be the set of heavy modules in $T$. Then, $\spr{\mathcal{H}}\leq k\br{T}$.
\begin{proof}
    Since $\mathcal{H}=\mathcal{H}_{\mathcal{T},c}\br{a}$, we have $\spr{\mathcal{H}}=k\br{\mathcal{T}, c, a}\leq \max_{t\in \mathbb{R}_{\geq 0}}k\br{T, c, t} = k\br{T, c}$.
\end{proof}
\end{observation}

\begin{observation}\label{subtreeKUpModularity}
    Let $T'$ be a subtree of $T$. Then, $k\br{T'}\leq k\br{T}$.
    \begin{proof}
        Fix any $t\in \mathbb{R}_{\geq 0}$ and let $H\in \mathcal{H}_{T, c}\br{t}$. We show that each such $H$ contributes at most 1 to $k\br{T', c, t}$. If $H\cap V\br{T'} = \emptyset$, then $H$ contributes 0. Otherwise, $H\cap V\br{T'}$ forms a connected subtree of $T'$, and thus contributes at most 1. The lemma follows.
    \end{proof}
\end{observation}

\begin{observation}\label{subtreePartialDt}
    Let $T'$ be a subtree of a tree $T$ and let $D'$ be a decision tree for $T'$. Then, $D'$ is a partial decision tree for $T$.
\end{observation}

\begin{observation}\label{neighborsPathObservation}
     Let $T'$ be a subtree of a tree $T$ and let $D$ be a partial decision tree for $T$ having no queries to vertices of $T'$, but containing at least one query to the vertices of $N_{T}\br{V\br{T'}}$. Let $Q$ denote the set of all such queries to vertices of $N_{T}\br{V\br{T'}}$ in $D$. Then, $D\angl{Q}$ forms a path in $D$.  
    \begin{proof}
        Let $q$ be any query in $D$. There are two cases:
        \begin{enumerate}
            \item $q\in V\br{T-V\br{T'}-N_{T}\br{V\br{T'}}}$. Then, for every $x\in N_{T}\br{V\br{T'}}$ being the target, $x$ belongs to the same connected component of $T-q$. Thus, no matter which vertex is the target, the answer is always the same. Therefore, $q$ has at most one child $u$ in $D$, such that $V\br{D_u}\cap Q \neq \emptyset$.
            \item $q\in N_{T}\br{V\br{T'}}$. After a query to $q$, the situation is as in the first case, except when $x=q$. Then, the response is $x$ itself, so no further queries are needed, and again $q$ has at most one child $u$ in $D$, such that $V\br{D_u}\cap Q \neq \emptyset$.
        \end{enumerate} 
        
        Since each $q\in Q$ has at most one child $u$ in $D$, with $D_u\cup Q\neq\emptyset$, $D\angl{Q}$ forms a path and the claim follows.
    \end{proof}
\end{observation}

\subsection{Base of the recursion}
We begin the description of our algorithm with the recursion base, which occurs whenever $b\leq{1}/{\log n}$ or for every $v\in V\br{\mathcal{T}}$, $c\br{v}>a$, i.e., every vertex is heavy. In such a situation, a solution is built by disregarding the costs of vertices and constructing a decision tree using the vertex ranking of $\mathcal{T}$. 

\begin{lemma}\label{baseOfRecursion}
    Let $D$ be a decision tree built, by calling \textsc{RankingBasedDT}$\br{\mathcal{T}}$ in line \ref{basecaseDT} of the \textsc{CreateDecisionTree} procedure. Then,  
$$
\COST_{D}\br{\mathcal{T}}\leq 2\cdot\OPT\br{T}.
$$
\begin{proof}
    There are two cases:
    \begin{enumerate}
        \item If $b\leq \frac{1}{\log n}$, then:
$$
\COST_{D}\br{\mathcal{T}}\leq\frac{\fl{\log n}+1}{\log n}\leq \frac{\log n+1}{\log n}\leq2\leq 2\cdot \OPT(\mathcal{T})\leq 2\cdot\OPT(T),
$$

where the first inequality is due to Corollary \ref{vertexRankingCorollary}, the fourth inequality follows from Observation \ref{basicBoundsOnCost}, and the last inequality is due to Observation \ref{subtreeOptLemma}.

        \item If for every $v\in V\br{\mathcal{T}}$, we have $c\br{v}> a$, then, define $c'\br{v}=a$ for all $v\in V\br{\mathcal{T}}$ (note that any value could be chosen here, since we treat each query as unitary). As $2c'\br{v} = 2a\geq b \geq c\br{v}$, we obtain $2\cdot \COST_D\br{\mathcal{T}, c'}\geq \COST_D\br{\mathcal{T}, c}$. Additionally, using the fact that $c'\br{v} \leq c\br{v}$, we have $\OPT\br{\mathcal{T}, c'}\leq \OPT\br{\mathcal{T}, c}$. Therefore:
$$
\COST_D\br{\mathcal{T}, c}\leq 2\cdot \COST_D\br{\mathcal{T}, c'}=2\cdot \OPT\br{\mathcal{T}, c'}\leq 2\cdot\OPT\br{\mathcal{T}, c}\leq 2\cdot\OPT\br{T, c},
$$

where the equality is due to Corollary \ref{vertexRankingCorollary} and the last inequality is due to Observation \ref{subtreeOptLemma}. The lemma follows.
    \end{enumerate}
\end{proof}
\end{lemma}

\begin{figure}[htp]
    \begin{minipage}[t]{0.45\textwidth}
    \centering
    \begin{tikzpicture}[every node/.style={draw, very thick}, every path/.style={very thick}]

    \node[circle, draw, minimum size = 0.75cm, fill=gray, drop shadow] (26) at (-3,-2) {};
    \node[circle, draw, minimum size=0.15cm, inner sep=0pt, fill=black] (27) at (-2.95,-1.95) {};
    \node[circle, draw, inner sep=0pt, fill=white] (41) at (-0.67,-2.433) {};
    \node[circle, draw, inner sep=0pt, fill=white] (42) at (-0.27,-2.38) {};
    \node[circle, draw, inner sep=0pt, fill=white] (43) at (-0.4,-1.57) {};
    
    \node[circle, draw, minimum size = 0.9cm, fill=gray, drop shadow] (28) at (-0.5,-2) {};
    \node[circle, draw, minimum size=0.15cm, inner sep=0pt, fill=black] (29) at (-0.5,-2) {};

    \node[circle, draw, minimum size=0.15cm, inner sep=0pt, fill=white] (30) at (1,-2.5) {};
    
    \node[circle, draw, minimum size = 0.75cm, fill=gray, drop shadow] (31) at (1.5,-3) {};
    \node[circle, draw, minimum size=0.15cm, inner sep=0pt, fill=gray!55] (32) at (1.4,-3) {};
    \node[circle, draw, inner sep=0pt, fill=white] (37) at (1.25,-3.3) {};
    \node[circle, draw, minimum size=0.15cm, inner sep=0pt, fill=black] (38) at (1.6,-2.79) {};
    \node[circle, draw, minimum size=0.15cm, inner sep=0pt, fill=white] (39) at (1.67,-3.025) {};
    
    \node[circle, draw, minimum size=0.15cm, inner sep=0pt, fill=white] (33) at (1.75,-3.5) {};
    
    \node[circle, draw, minimum size = 0.75cm, fill=gray, drop shadow] (34) at (1.9,-4) {};
    \node[circle, draw, minimum size=0.15cm, inner sep=0pt, fill=black] (35) at (2.05,-3.95) {};
    
    \node[circle, draw, minimum size = 1cm, fill=gray, drop shadow] (1) at (-3,2.75) {};
    \node[circle, draw, minimum size=0.15cm, inner sep=0pt, fill=black] (2) at (-3,3) {};

    \node[circle, draw, minimum size = 1.35cm, fill=gray, drop shadow] (3) at (-1,2.25) {};
    \node[circle, draw, minimum size=0.15cm, inner sep=0pt, fill=black] (4) at (-0.85,2.5) {};
    \node[circle, draw, inner sep=0pt, fill=white] (36) at (-0.3,2.25) {};
    \node[circle, draw, inner sep=0pt, fill=white] (40) at (-1.35,1.66) {};

    \node[circle, draw, minimum size=0.15cm, inner sep=0pt, fill=white] (5) at (-2.77,2) {};
    
    \node[circle, draw, minimum size=0.15cm, inner sep=0pt, fill=white] (6) at (-0.7,1.25) {};
    
    \node[circle, draw, minimum size = 0.75cm, fill=gray, drop shadow] (7) at (1.7,0.9) {};
    \node[circle, draw, minimum size=0.15cm, inner sep=0pt, fill=black] (8) at (1.75,0.95) {};
    
    \node[circle, draw, inner sep=0pt, fill=white] (9) at (-2.53,1) {};
    
    \node[circle, draw, minimum size=0.15cm, inner sep=0pt, fill=white] (10) at (1,0.65) {};
    
    \node[circle, draw, inner sep=0pt, fill=white] (11) at (0.5,0.47) {};
    
    \node[circle, draw, minimum size=0.15cm, inner sep=0pt, fill=gray!55] (12) at (-0.52,0.1) {};
    
    \node[circle, draw, minimum size=0.15cm, inner sep=0pt, fill=white] (13) at (-0.95,-0.1) {};

    \node[circle, draw, minimum size = 1cm, fill=gray, drop shadow] (14) at (-2,-0.5) {};
    \node[circle, draw, minimum size=0.15cm, inner sep=0pt, fill=black] (15) at (-2.2,-0.4) {};
    \node[circle, draw, minimum size=0.15cm, inner sep=0pt, fill=white] (16) at (-2,-0.45) {};
    \node[circle, draw, minimum size=0.15cm, inner sep=0pt, fill=gray!55] (17) at (-1.8,-0.5) {};
    
    \node[circle, draw, minimum size=0.15cm, inner sep=0pt, fill=white] (18) at (-1.35,-1.1) {};

    \node[circle, draw, minimum size=0.15cm, inner sep=0pt, fill=white] (19) at (-2.4,-1.25) {};
    
    \node[circle, draw, inner sep=0pt, fill=white] (20) at (-1.05,-1.45) {};
    
    \node[circle, draw, minimum size = 0.6cm, fill=gray, drop shadow] (21) at (1.5,-1.55) {};
    \node[circle, draw, minimum size=0.15cm, inner sep=0pt, fill=black] (22) at (1.5,-1.55) {};
    
    \node[circle, draw, minimum size=0.15cm, inner sep=0pt, fill=white] (23) at (1,-1.65) {};
    
    \node[circle, draw, minimum size=0.15cm, inner sep=0pt, fill=gray!55] (24) at (0.5,-1.775) {};
    
    \node[circle, draw, minimum size=0.15cm, inner sep=0pt, fill=white] (25) at (0.25,-1.85) {};

    \draw[] (2) to (5);
    \draw[] (4) to (6);
    \draw[] (5) to (9);
    \draw[] (6) to (12);
    \draw[] (8) to (10);
    \draw[] (9) to (15);
    \draw[] (10) to (11);
    \draw[] (11) to (12);
    \draw[] (12) to (13);
    \draw[] (13) to (17);
    \draw[] (15) to (16);
    \draw[] (16) to (17);
    \draw[] (17) to (18);
    \draw[] (17) to (19);
    \draw[] (18) to (20);
    \draw[] (19) to (27);
    \draw[] (20) to (29);
    \draw[] (22) to (23);
    \draw[] (23) to (24);
    \draw[] (24) to (25);
    \draw[] (24) to (30);
    \draw[] (25) to (29);
    \draw[] (30) to (32);
    \draw[] (32) to (33);
    \draw[] (32) to (39);
    \draw[] (33) to (35);
    \draw[] (38) to (39);
    \draw[very thick, fill=gray!20, drop shadow]
    (9).. controls +(-120:1.5cm) and +(-200:1.5cm).. (9);
    \draw[very thick, fill=gray!20, drop shadow]
    (11).. controls +(-30:3cm) and +(-110:3cm).. (11);
    \draw[very thick, fill=gray!20, drop shadow]
    (20).. controls +(-105:2cm) and +(-185:2cm).. (20);
    \draw[very thick, fill=gray!20, drop shadow]
    (36).. controls +(30:3.6cm) and +(-50:3.6cm).. (36);
    \draw[very thick, fill=gray!20, drop shadow]
    (37).. controls +(-80:2cm) and +(-160:2cm).. (37);
    \draw[very thick, fill=gray!20, drop shadow]
    (40).. controls +(-80:1.75cm) and +(-160:1.75cm).. (40);
    \draw[very thick, fill=gray!20, drop shadow]
    (41).. controls +(-70:3cm) and +(-150:3cm).. (41);
    \draw[very thick, fill=gray!20, drop shadow]
    (42).. controls +(-20:1.5cm) and +(-100:1.5cm).. (42);
    \draw[very thick, fill=gray!20, drop shadow]
    (43).. controls +(120:1cm) and +(30:1cm).. (43);
    
    \end{tikzpicture}
    \caption{Example tree $\mathcal{T}$. Dark grey circles represent heavy modules. Light grey regions represent light subtrees. Black vertices represent $\mathcal{X}$. Gray and black vertices represent $\mathcal{Y}$. White, gray and black vertices represent $\mathcal{Z}$. Lines represent paths of vertices between vertices of $\mathcal{Z}$.}\label{exampleTreeWithSetZ}
\end{minipage}
    \hspace{0.06\textwidth} 
    \begin{minipage}[t]{0.45\textwidth}
    \centering
    \begin{tikzpicture}[every node/.style={draw, very thick, drop shadow}, every path/.style={very thick}]
    
    \node[circle, draw, minimum size=0.3cm, inner sep=0pt, fill=black] (2) at (-3,4) {};

    \node[circle, draw, minimum size=0.3cm, inner sep=0pt, fill=black] (4) at (-0.85,3.5) {};

    \node[circle, draw, minimum size=0.3cm, inner sep=0pt, fill=white] (5) at (-2.77,3) {};
    
    \node[circle, draw, minimum size=0.3cm, inner sep=0pt, fill=white] (6) at (-0.7,2.25) {};
    
    \node[circle, draw, minimum size=0.3cm, inner sep=0pt, fill=black] (8) at (2,2) {};

    \node[circle, draw, minimum size=0.3cm, inner sep=0pt, fill=white] (10) at (1,1.65) {};

    \node[circle, draw, minimum size=0.3cm, inner sep=0pt, fill=gray!55] (12) at (-0.52,1.1) {};
    
    \node[circle, draw, minimum size=0.3cm, inner sep=0pt, fill=white] (13) at (-0.95,0.9) {};

    \node[circle, draw, minimum size=0.3cm, inner sep=0pt, fill=black] (15) at (-2.5,0.6) {};
    \node[circle, draw, minimum size=0.3cm, inner sep=0pt, fill=white] (16) at (-2,0.55) {};
    \node[circle, draw, minimum size=0.3cm, inner sep=0pt, fill=gray!55] (17) at (-1.5,0.5) {};
    
    \node[circle, draw, minimum size=0.3cm, inner sep=0pt, fill=white] (18) at (-1.35,-0.1) {};

    \node[circle, draw, minimum size=0.3cm, inner sep=0pt, fill=white] (19) at (-2.4,-0.25) {};

    \node[circle, draw, minimum size=0.3cm, inner sep=0pt, fill=black] (22) at (1.5,-0.55) {};
    
    \node[circle, draw, minimum size=0.3cm, inner sep=0pt, fill=white] (23) at (1,-0.65) {};
    
    \node[circle, draw, minimum size=0.3cm, inner sep=0pt, fill=gray!55] (24) at (0.5,-0.775) {};
    
    \node[circle, draw, minimum size=0.3cm, inner sep=0pt, fill=white] (25) at (0,-0.85) {};

    \node[circle, draw, minimum size=0.3cm, inner sep=0pt, fill=black] (27) at (-2.95,-0.95) {};
    
    \node[circle, draw, minimum size=0.3cm, inner sep=0pt, fill=black] (29) at (-0.5,-1) {};

    \node[circle, draw, minimum size=0.3cm, inner sep=0pt, fill=white] (30) at (1,-1.5) {};
    
    \node[circle, draw, minimum size=0.3cm, inner sep=0pt, fill=gray!55] (32) at (1.4,-2) {};
    \node[circle, draw, minimum size=0.3cm, inner sep=0pt, fill=black] (38) at (1.9,-1.55) {};
    \node[circle, draw, minimum size=0.3cm, inner sep=0pt, fill=white] (39) at (1.8,-1.95) {};
    
    \node[circle, draw, minimum size=0.3cm, inner sep=0pt, fill=white] (33) at (2,-2.5) {};
    
    \node[circle, draw, minimum size=0.3cm, inner sep=0pt, fill=black] (35) at (2.55,-2.95) {};
    
    \draw[] (2) to (5);
    \draw[] (4) to (6);
    \draw[] (5) to (15);
    \draw[] (6) to (12);
    \draw[] (8) to (10);
    \draw[] (10) to (12);
    \draw[] (12) to (13);
    \draw[] (13) to (17);
    \draw[] (15) to (16);
    \draw[] (16) to (17);
    \draw[] (17) to (18);
    \draw[] (17) to (19);
    \draw[] (18) to (29);
    \draw[] (19) to (27);
    \draw[] (22) to (23);
    \draw[] (23) to (24);
    \draw[] (24) to (25);
    \draw[] (24) to (30);
    \draw[] (25) to (29);
    \draw[] (30) to (32);
    \draw[] (32) to (33);
    \draw[] (32) to (39);
    \draw[] (33) to (35);
    \draw[] (38) to (39);
    \end{tikzpicture}
    \caption{Auxiliary tree $\mathcal{T}_{\mathcal{Z}}$ built from vertices of set $\mathcal{Z}$.}\label{exampleAuxTreeTZ}
\end{minipage}
\begin{minipage}[t]{1\textwidth}
\centering
    \begin{tikzpicture}[scale=1.1]
    \draw[very thick, fill=black!80, drop shadow] 
    (4,-4) -- (4.9,-5.8) -- (3.1,-5.8) -- cycle; 
    
    \draw[very thick, fill=gray, drop shadow] (2.5,-6.5) -- (3.2,-7.9) -- (1.8,-7.9) -- cycle; 
    
    \draw[very thick, fill=gray, drop shadow] (3.8,-6.4) -- (4.3,-7.4) -- (3.3,-7.4) -- cycle; 
    
    \draw[very thick, fill=gray!20, drop shadow] (4.8,-6.3) -- (5.2,-7.1) -- (4.4,-7.1) -- cycle;
    
    \node at (5.75, -7) {$\dots$}; 
    
    \draw[very thick, fill=gray, drop shadow] (6.5,-6.5) -- (7.2,-8.1) -- (5.8,-8.1) -- cycle;

    \draw[very thick] (3.1,-5.8) -- (2.5,-6.5);
    
    \draw[very thick] (3.6,-5.8) -- (3.8,-6.4); 
    
    \draw[very thick] (4.2,-5.8) -- (4.8,-6.3); 
    
    \draw[very thick] (4.9,-5.8) -- (6.5,-6.5); 

    \draw[very thick, fill=gray!20, drop shadow] (1.2,-8.5) -- (1.8,-9.7) -- (0.6,-9.7) -- cycle;

    \draw[very thick, fill=gray!20, drop shadow] (2.6,-8.4) -- (3.1,-9.4) -- (2.1,-9.4) -- cycle;
    
    \node at (3.25, -8.7) {$\dots$}; 

    \draw[very thick, fill=gray!20, drop shadow] (3.9,-8.3) -- (4.3,-9.1) -- (3.5,-9.1) -- cycle;
    
    \draw[very thick] (1.8,-7.9) -- (1.2,-8.5);
    \draw[very thick] (2.5,-7.9) -- (2.6,-8.4);
    \draw[very thick] (3.2,-7.9) -- (3.9,-8.3);

    \draw[very thick, fill=gray!20, drop shadow] (5,-8.8) -- (5.7,-10.2) -- (4.3,-10.2) -- cycle;

    \draw[very thick, fill=gray!20, drop shadow] (6.5,-8.7) -- (7.1,-9.9) -- (5.9,-9.9) -- cycle;
    
    \node at (7.2, -9) {$\dots$}; 

    \draw[very thick, fill=gray!20, drop shadow] (7.9,-8.6) -- (7.4,-9.6) -- (8.4,-9.6) -- cycle;
    
    \draw[very thick] (5.8,-8.1) -- (5,-8.8);
    \draw[very thick] (6.3,-8.1) -- (6.5,-8.7);
    \draw[very thick] (7.2,-8.1) -- (7.9,-8.6);

    \draw [decorate, 
       decoration = {calligraphic brace, amplitude = 10pt, mirror}, 
       line width = 1.5pt] 
      (0.5,-3.85) -- (0.5,-10.25);

    \node at (-0.6, -7.05) {$\COST_{D}\br{\mathcal{T}}$};
        
    \end{tikzpicture}
    \caption{The structure of the decision tree $D$, built by the Algorithm \ref{createDecisionTree}. The dark gray subtree represents the decision tree $D_{\mathcal{Z}}$, obtained by calling the QPTAS for  $\mathcal{T}_{\mathcal{Z}}$, $c$ and $\epsilon=1$. Gray subtrees represent decision trees $D_H$, each built for a unique heavy module $H\subseteq V\br{\mathcal{T}'}$ of every $\mathcal{T}'\in \mathcal{T}-\mathcal{Z}$, by calling the \textsc{RankingBasedDT} procedure for $\mathcal{T}'\angl{H}$. Light gray subtrees represent decision trees $D_L$, built for each $L\in \mathcal{T}'-H$, by recursively calling \textsc{CreateDecisionTree} with $L$, $c$ and $\left(a/2,a\right]$.}\label{structure_of_dt}
\end{minipage}
\end{figure}

\subsection{Construction of the Auxiliary Tree}\label{auxTreeConstruction}

To obtain the solution for the non-base case of our algorithm, we first construct the so-called \textit{auxiliary tree}. To do so, we begin by defining a set $\mathcal{X}\subseteq V\br{\mathcal{T}}$. For every heavy module $H\in\mathcal{H}$, we pick an arbitrary $v\in H$ and add it to $\mathcal{X}$. We also define a set 
$\mathcal{Y}=\mathcal{X}\cup\brc{v\in V\br{\mathcal{T}\angl{\mathcal{X}}} | \deg_{\mathcal{T}\angl{\mathcal{X}}}\br{v}\geq 3}$, by extending $\mathcal{X}$ to contain all vertices with degree at least $3$ in $T\angl{\mathcal{X}}$. Furthermore, we define a set $\mathcal{Z}\subseteq V\br{\mathcal{T}}$ consisting of the vertices in $\mathcal{Y}$ and, for every $u,v\in \mathcal{Y}$, such that $\mathcal{P}_{\mathcal{T}}\br{u, v}\neq\emptyset$ and $\mathcal{P}_{\mathcal{T}}\br{u, v}\cap \mathcal{Y}=\emptyset$, we add to $\mathcal{Z}$ the lightest vertex between them, i. e., 
$v_{u,v} = \argmin_{z\in \mathcal{P}_{\mathcal{T}}\br{u, v}}\brc{c\br{z}}$.
To see an example of construction of the sets $\mathcal{X}, \mathcal{Y}, \mathcal{Z}$, see Figure \ref{exampleTreeWithSetZ}. 

We then create the auxiliary tree 
$\mathcal{T}_{\mathcal{Z}}=\br{\mathcal{Z},\brc{uv | \mathcal{P}_{\mathcal{T}}\br{u, v}\cap \mathcal{Z}=\emptyset}}$ 
(for an example, see Figure \ref{exampleAuxTreeTZ}). Our algorithm starts by building a decision tree $D_{\mathcal{Z}}$ for $\mathcal{T}_{\mathcal{Z}}$, by taking $\epsilon=1$ and applying the QPTAS from Theorem \ref{QPTAS}. Observe that, since $D_{\mathcal{Z}}$ is a partial decision tree for $\mathcal{T}$ and corresponding vertices in $\mathcal{T}$ and $\mathcal{T}_{\mathcal{Z}}$ have the same costs, we have that:

\begin{observation}\label{CostDZinTObservation}
    $\COST_{D_{\mathcal{Z}}}\br{\mathcal{T}_{\mathcal{Z}}}=\COST_{D_{\mathcal{Z}}}\br{\mathcal{T}}$.
\end{observation}

Let $D = D_{\mathcal{Z}}$. For each connected component $\mathcal{T}'\in \mathcal{T}-\mathcal{Z}$, we build a new decision tree as follows: By the construction of $\mathcal{Z}$, all heavy vertices in $V\br{\mathcal{T}'}$ form a single heavy module $H\subseteq V\br{\mathcal{T}'}$. We create a new decision tree $D_H$ for $\mathcal{T}'\angl{H}$, by calling the \textsc{RankingBasedDT} procedure with argument $\mathcal{T}'\angl{H}$ and we hang $D_{H}$ in $D$ below the unique last query to a vertex in $N_\mathcal{T}\br{\mathcal{T}'}$ (which is possible due to Observation \ref{neighborsPathObservation}). As, by Observation \ref{subtreePartialDt}, $D_H$ is a partial decision tree for $\mathcal{T}'$, it follows that $D$ is also a partial decision tree for $\mathcal{T}$. 

Now notice that for each $L\in \mathcal{T}'-H$, there is no $v\in V\br{L}$, such that $c\br{v}>a$. This allows us to create a decision tree $D_L$ recursively, by calling the \textsc{CreateDecisionTree} procedure with arguments $L$, $c$ and $\left(a/2,a\right]$. Next, we hang $D_L$ in $D$ below the unique last query to a vertex in $N_{\mathcal{T}'}\br{L}$ (again, using Observation \ref{neighborsPathObservation}). Since after all such operations, every vertex $v\in V\br{\mathcal{T}}$ also belongs to $D$, we obtain a valid decision tree $D$ for $\mathcal{T}$. To see example structure of such solution, see Figure \ref{structure_of_dt}.

\subsection{Analysis of the algorithm}\begin{lemma}\label{auxTreeSizeLemma}
    Let $\mathcal{T}_{\mathcal{Z}}$ be the auxiliary tree. Then, $\spr{V\br{\mathcal{T}_{\mathcal{Z}}}}\leq 4k-3$.
    \begin{proof}
        First, we show that $\spr{\mathcal{Y}}\leq 2k-1$. We use induction on the elements of $\mathcal{H}$. We construct a family of sets $\mathcal{H}_1, \mathcal{H}_2, \dots, \mathcal{H}_{\spr{\mathcal{H}}}$, such that for every integer $1\leq h \leq \spr{\mathcal{H}}$, $\spr{\mathcal{H}_h}=h$ and $\mathcal{H}_{\spr{\mathcal{H}}} = \mathcal{H}$. For each $\mathcal{H}_h$, we also construct a corresponding set $\mathcal{Y}_h$, eventually ensuring that $\mathcal{Y}_{\spr{\mathcal{H}}}=\mathcal{Y}$.
        
        Let $\mathcal{H}_1=\emptyset$, $\mathcal{Y}_1=\emptyset$. Pick any heavy module $H\subseteq V\br{\mathcal{T}}$ and add it to $\mathcal{H}_1$. Add the unique vertex $v$, such that $v\in H\cap\mathcal{X}$ to $\mathcal{Y}_1$, so that $\spr{\mathcal{Y}_1}=1$. Assume by induction that for some $h\geq 1$, $\spr{\mathcal{Y}_h}\leq 2h-1$.  
        Two heavy modules $H_1,H_2\subseteq V\br{\mathcal{T}}$ will be called  \textit{neighbors} if for every $H_3\subseteq V\br{\mathcal{T}}$ with $H_3\neq H_1,H_2$, we have $\mathcal{P}_{\mathcal{T}}\br{H_1,H_2}\cap H_3 = \emptyset$. Pick $H\in\mathcal{H}$, such that $H\notin \mathcal{H}_h$ to be a heavy module that is a neighbor of some member of $\mathcal{H}_h$. We define $\mathcal{H}_{h+1}=\mathcal{H}_h\cup\brc{H}$. 
        Let $z$ be the unique vertex, such that $v\in H\cap\mathcal{X}$, and let $\mathcal{Y}_{h+1}=\mathcal{Y}_h \cup \brc{z}$. Define $\mathcal{T}_{h+1} = \mathcal{T}\angl{\brc{v\in \mathcal{Y}_{h+1} | \mathcal{P}_\mathcal{T}\br{v,z} \cap \mathcal{Y}_{h+1} = \emptyset}}$. Note that $\mathcal{T}_{h+1}$ is a spider (a tree with at most one vertex of degree above 2). Add to $\mathcal{Y}_{h+1}$ the unique vertex $v\in V\br{\mathcal{T}_{h+1}}$, such that $\deg_{\mathcal{T}_{h+1}}\br{v}\geq 3$, if it exists. Clearly, $\spr{\mathcal{Y}_{h+1}} \leq 2h+1$, completing the induction.  

        By construction, $\mathcal{H}_{\spr{\mathcal{H}}}=\mathcal{H}$ and $\mathcal{Y}_{\spr{\mathcal{H}}}=\mathcal{Y}$, so $\spr{\mathcal{Y}} \leq 2\cdot \spr{\mathcal{H}} - 1 \leq 2k-1$ where the last inequality is by Observation \ref{heavymodulesetsize}. As paths between vertices in $\mathcal{Y}$ form a tree when contracted, at most $2k-2$ additional vertices are added while constructing $\mathcal{Z}$ (at most one per path). The lemma follows.
    \end{proof}
\end{lemma}

\begin{lemma}\label{auxTreeCostLemma}
    Let $\mathcal{T}_{\mathcal{Z}}$ be the auxiliary tree. Then, $\OPT\br{\mathcal{T}_{\mathcal{Z}}}\leq \OPT\br{\mathcal{T}}$.
    \begin{proof}
        Let $D^*$ be the decision tree for $\mathcal{T}\angl{\mathcal{Z}}$. We build a new decision tree $D_{\mathcal{Z}}'$ for $\mathcal{T}_{\mathcal{Z}}$ by transforming $D^*$ as follows: 
        
        Let $u,v\in \mathcal{Y}$, such that $\mathcal{P}_{\mathcal{T}}\br{u, v}\neq\emptyset$ and $\mathcal{P}_{\mathcal{T}}\br{u, v}\cap \mathcal{Y}=\emptyset$. Let $q\in V\br{D^*}$ be the first query to a vertex among $\mathcal{P}_{\mathcal{T}}\br{u, v}$. Recall that we picked $v_{u,v}=\argmin_{z\in \mathcal{P}_{\mathcal{T}}\br{u, v}}\brc{c\br{z}}$, so $c\br{v_{u,v}}\leq c\br{q}$. 
        We replace $q$ in $D^*$ with the query to $v_{u,v}$ and delete all queries to vertices in $\mathcal{P}_{\mathcal{T}}\br{u, v}-v_{u,v}$. By construction, $D_{\mathcal{Z}}'$ is a valid decision tree for $\mathcal{T}_{\mathcal{Z}}$, and by choosing $v_{u,v}$ to minimize $c$, we did not increase the cost, so we have that: 
        $$
        \COST_{D_{\mathcal{Z}}'}\br{\mathcal{T}_{\mathcal{Z}}} \leq \OPT\br{\mathcal{T}\angl{\mathcal{Z}}}.
        $$
        
        Therefore, we have:
        $$
        \OPT\br{\mathcal{T}_{\mathcal{Z}}}\leq \COST_{D_{\mathcal{Z}}'}\br{\mathcal{T}_{\mathcal{Z}}}\leq \OPT\br{\mathcal{T}\angl{\mathcal{Z}}}\leq \OPT\br{\mathcal{T}},
        $$
        
        where the first inequality is due to the definition of optimality and the last inequality follows by Lemma \ref{subtreeOptLemma}.
    \end{proof}
\end{lemma}

\begin{lemma}\label{heavygroupcostlemma}
    Let $H$ be the unique heavy module of $\mathcal{T}'\in \mathcal{T}-\mathcal{Z}$. Then, the decision tree $D_H$ is of cost at most:
    $$\COST_{D_H}\br{\mathcal{T}'\angl{H}}\leq 2\cdot\OPT\br{\mathcal{T}}.$$
    \begin{proof}
        
    For every $v\in H$ let $c'\br{v}=a$. We have $2 c'\br{v}\geq b c'\br{v}/a =  b \geq c\br{v}$ so we get that $2\cdot\COST_{D_H}\br{\mathcal{T}'\angl{H}, c'}\geq \COST_{D_H}\br{\mathcal{T}'\angl{H}, c}$. Additionally, using the fact that $c'\br{v} \leq c\br{v}$ we have that $\OPT\br{\mathcal{T}'\angl{H}, c'}\leq \OPT\br{\mathcal{T}'\angl{H}, c}$. Hence:
    
    \begin{align*}
        \COST_{D_H}\br{\mathcal{T}'\angl{H}, c}&\leq 2\cdot\COST_{D_H}\br{\mathcal{T}'\angl{H}, c'}=2\cdot\OPT\br{\mathcal{T}'\angl{H}, c'} \\
        &\leq
        2\cdot\OPT\br{\mathcal{T}'\angl{H}, c}\leq 2\cdot\OPT\br{\mathcal{T}, c}
    \end{align*}

    where the equality is by the Corollary \ref{vertexRankingCorollary} and the last inequality is due to the fact that $\mathcal{T}'\angl{H}$ is a subtree of $\mathcal{T}'$, which is a subtree of $\mathcal{T}$ (Lemma \ref{subtreeOptLemma}).
    \end{proof}
\end{lemma}

\subsection{The main result}

Let $d$ be the remaining depth of the recursion call performed in Line~\ref{recursion} of the algorithm, i.e., the number of recursive steps from the current call to the base case (for the base case, this value is equal to $d=0$). We show that at each level of the recursion we pay $O\br{\OPT\br{T}}$, so the approximation ratio of the algorithm is bounded by $O\br{d}$:
\begin{lemma}\label{costofthesolution}
$\COST_D\br{\mathcal{T}}\leq\br{4 d+2}\cdot\OPT\br{T}$.
    \begin{proof}
        
    Let $Q_D\br{\mathcal{T},x}$ be the sequence of queries performed in order to find $x\in V\br{\mathcal{T}}$. By construction of the Algorithm \ref{createDecisionTree}, $Q_D\br{\mathcal{T},x}$ consists of at most three distinct subsequences of queries (see Figure \ref{structure_of_dt}):
    \begin{enumerate}
        \item Firstly, there is a sequence of queries belonging to $Q_{D_{\mathcal{Z}}}\br{\mathcal{T}_{\mathcal{Z}},x}$.
        \item If $x\notin \mathcal{Z}$, then, there is a sequence of queries belonging to $Q_{D_{H}}\br{\mathcal{T}'\angl{H},x}$ for a unique heavy group $H\subseteq V\br{\mathcal{T}'}$ of $\mathcal{T}'\in\mathcal{T}-\mathcal{Z}$, such that $x\in \mathcal{T}'$.
        \item At last, if $x\notin H$, there is a sequence of queries belonging to $Q_{D_{L}}\br{L,x}$ for $L\in \mathcal{T}'-H$, such that $x\in V\br{L}$. 
    \end{enumerate}  

    Note that it sometimes may happen that some of the above sequences are empty.
    
    We prove by induction that $\COST_{D}\br{\mathcal{T}}\leq \br{4d +2}\cdot \OPT\br{T}$. When $d=0$ (the base case), the induction hypothesis is true, due to the Lemma \ref{baseOfRecursion}. For $d>0$, assume by induction that the cost of the decision tree  built for each $L$, is at most $\COST_{D_L}\br{L}\leq \br{4\br{d-1} +2}\cdot \OPT\br{T}$. We have:
        \begin{align*}
        \COST_D\br{\mathcal{T}}
        &\leq
        \COST_{D_{\mathcal{Z}}}\br{\mathcal{T}}+\max_{\mathcal{T}'\in \mathcal{T}-\mathcal{Z}}\brc{\COST_{D_H}\br{\mathcal{T}'\angl{H}}+\max_{L\in \mathcal{T}'-H}\brc{\COST_{D_L}\br{L}}}
         \\ 
        &\leq
        \COST_{D_{\mathcal{Z}}}\br{\mathcal{T}_{\mathcal{Z}}}+\max_{\mathcal{T}'\in \mathcal{T}-\mathcal{Z}}\brc{2\cdot\OPT\br{\mathcal{T}}+\br{4\br{d-1} +2}\cdot \OPT\br{T}}
        \\ 
        &\leq 
        2\cdot\OPT\br{\mathcal{T}_{\mathcal{Z}}}+2\cdot\OPT\br{T}+ \br{4\br{d-1} +2}\cdot \OPT\br{T} 
        \\
        &
        \leq 
        2\cdot\OPT\br{\mathcal{T}}+4d\cdot \OPT\br{T} = \br{4d+2}\cdot \OPT\br{T}
        \end{align*}
        
        where the first inequality is due to the construction of the decision tree returned by the Algorithm \ref{createDecisionTree}, the second inequality is by Observation \ref{CostDZinTObservation}, Observation \ref{heavygroupcostlemma} and by the induction hypothesis, the third inequality is due to Theorem \ref{QPTAS} and using the fact that $\mathcal{T}$ is a subtree of $T$ (Lemma \ref{subtreeOptLemma}) and the last inequality is due to Lemma \ref{auxTreeCostLemma}.
        
    \end{proof}
\end{lemma}

We are now ready to prove our main theorem: 
\begin{theorem}
\label{parametrizedAlgorithm}
    There exists an $O\br{\log\log n}$-approximation algorithm for the Tree Search Problem running in $k^{O\br{\log k}}\cdot\text{poly}\br{n}$ time.
    \begin{proof}
        Let $D = \textsc{CreateDecisionTree}\br{T, \left({2^{\cl{\log\log n}-1}}/{\log n},1\right]}$. Since there are at most $\cl{\log\log n}+1$ intervals processed, the depth of the recursion is bounded by 
        $
        d \leq \cl{\log\log n} \leq \log\log n + 1
        $.
        Hence, using Lemma \ref{costofthesolution} we get that:
        $$
        \COST_{D}\br{T} \leq (4 \cdot \log\log n + 6) \cdot \OPT\br{T} = O\br{\log\log n \cdot \OPT\br{T}}.
        $$

        By Observation \ref{subtreeKUpModularity}, for every subtree $\mathcal{T}$ of $T$, processed at some level of the recursion, we have $k\br{\mathcal{T}} \leq k\br{T}$. Using Lemma \ref{auxTreeSizeLemma}, at each such level the call to the QPTAS from Theorem \ref{QPTAS} (line \ref{QPTAScall} of the \textsc{CreateDecisionTree}) runs in time bounded by: 
        $$
        k\br{\mathcal{T}}^{O\br{\log (4 \cdot k\br{\mathcal{T}})}} = k\br{T}^{O\br{\log k\br{T}}}.
        $$
        
        Since $d = O(\text{poly}(n))$ and all other computation can be performed in polynomial time, the overall running time is bounded by 
        $
        k^{O\br{\log k}} \cdot \text{poly}\br{n}
        $,
        as required.
    \end{proof}
\end{theorem}

\section{Conclusions and future work}\label{conslusionsAndFutureWork}

We have shown that approximating the Tree Search Problem within a factor of $O\br{\log\log n}$ is FPT with respect to the $k$-up-modularity of $c$ in $T$. Additionally, the running time is quasipolynomial in $k$, making the algorithm practical. It should be noted that if desired, one can also replace the usage of QPTAS with the polynomial-time $O\br{\sqrt{\log n}}$-approximation algorithm to obtain a polynomial-time $O\br{\sqrt{\log k}\log\log n}$-approximation algorithm.

The approximation factor achieved by our algorithm is significantly better than the state-of-the-art $O\br{\sqrt{\log n}}$ known for general cost functions \cite{dereniowski2017ApproxSsForGeneralBSinWTs}. Although a constant-factor approximation is achievable for $k=1$ \cite{dereniowski2022CFApproxAlgForBSInTsWithMonoQTimes, dereniowski2024SInTsMonoQTs}, such a solution exploits the strictly monotonic nature of the cost function and is similar in structure to the much simpler, uniform-cost version of the problem. In practical applications, one should not expect perfectly monotonic cost functions. In such cases, our solution may be substantially more useful than reverting to the general $O\br{\sqrt{\log n}}$-approximation algorithm.

Many seemingly complex instances with high $k$-up modularity can be transformed into instances with low $k$ without introducing excessive error. For example, consider a tree in which the cost function alternates between $1$ and $1+\epsilon$ along each path. For such instances it is possible to have $k = O\br{n}$. However, by rounding all costs up to the nearest multiple of a some constant (e.g., $2$), one obtains an instance with $k=1$. In general, such operation increases the optimal cost by at most a constant factor and ensures the stability of our algorithm under small cost perturbations.

The central question in our line of research is whether there exists a polynomial-time constant-factor approximation algorithm (or ideally a PTAS) for an arbitrary tree $T$ and cost function $c$. We believe that the approach used to obtain the
current state-of-the-art $O\br{\sqrt{\log n}}$-approximate algorithm \cite{dereniowski2017ApproxSsForGeneralBSinWTs}, based on recursive processing of
the tree's center, is unlikely to lead to further improvement of the approximation ratio. In contrast, an approach similar to ours, based on recursive processing of cost intervals, could potentially reduce the approximation factor closer to $O\br{\log \log n}$.

On the hardness side, it remains an open question whether inapproximability bounds can be established for this problem. Currently, the only known result is the trivial nonexistence of a constant-additive-factor approximation algorithm. A promising direction for future research is to derive stronger, more sophisticated inapproximability bounds.




\bibliography{lipics-v2021-sample-article}

\end{document}